 \documentclass[10pt]{iopart}

 \usepackage{graphicx}
\begin{document}

 \def\BE{\begin{equation}}
 \def\EE{\end{equation}}
 \def\BA{\begin{array}}
 \def\EA{\end{array}}
 \def\BEA{\begin{eqnarray}}
 \def\EEA{\end{eqnarray}}
 \def\nn{\nonumber}
 \def\ra{\rangle}
 \def\la{\langle}
 \def\E{{\cal E}}
 \def\tE{\tilde{\cal E}}
 \def\g{\gamma_\perp}
 \def\k{\kappa}
 \def\t{\tau}
 \def\T{{\cal T}}

 \title[Cavity-assisted squeezing and entanglement:
 Non-adiabatic effects]{Cavity-assisted squeezing and entanglement:
 Non-adiabatic effects and optimal cavity--atomic ensemble matching}

 \author{N~I~Masalaeva$^{(1)}$, A~N~Vetlugin$^{(2)}$, and I~V~Sokolov$^{(1)}$}
 \address{1 St. Petersburg State University,  7/9 Universitetskaya nab., St. Petersburg,
 199034 Russia}
 \address{2 Centre for Disruptive Photonic Technologies, SPMS, TPI, Nanyang Technological University,
 637371  Singapore}
 %\date{}
 \ead{i.sokolov@mail.spbu.ru, sokolov.i.v@gmail.com}

 \begin{abstract}

We investigate theoretically quantum entanglement of light with the collective spin polarization of a
cold atomic ensemble in cavity-assisted Raman schemes.    Previous works concentrated mostly on the bad cavity limit
where the signals are much longer than the cavity field lifetime. In view of atomic relaxation and other imperfections,
there may arise a need to speed-up the light-atoms interface operation. By increasing the cavity field lifetime, one can
achieve better light-matter coupling and entanglement. In our work, we consider the non-adiabatic effects that become
important beyond the bad cavity limit in both low-photon and continuous variables regime.
We find classical control field time profiles that allow one to retrieve from the cavity an output quantized
signal of a predefined time shape and duration, which is optimal for the homodyne detection, optical mixing or
further manipulation. This is done  for a wide range of the signal duration as compared to the cavity field lifetime.
We discuss an optimal cavity--atomic ensemble matching in terms of the cavity field lifetime which allows one to apply
less intense control field and  to minimize a variety of non-linear effects, such as  AC light shifts,
four-wave mixing, etc, which may be potentially harmful to an experiment.\\ \\
Keywords: Quantum entanglement, cavity QED, non-linear light-matter interaction

 \end{abstract}

 \pacs{42.50.Ex, 32.80.Qk }

 %\submitto{Physica Scripta}

 \maketitle

 %\tableofcontents

 \section{Introduction}

Quantum entanglement is an important resource for quantum information, including high-speed and
long-distance quantum communication, circuit quantum computation etc.
Raman scattering \cite{Raymer90} is commonly used to produce the light-matter entanglement.
At an early stage in the study of statistical properties of  Raman scattering, the main focus was on the
statistics of Stokes light itself. An important contribution to the theory of fluctuations in superfluorescence,
including the Raman superfluorescence, was due to Glauber et al. \cite{Glauber79}.

In order to create light-matter entanglement at single-quantum level, pump field photon is exchanged for the
Stokes or anti-Stokes photon and matter excitation.
A single photon generation \cite{Kuhn15} was achieved with a single atom, trapped ion, color center in a solid, and quantum
dot. By using an atomic ensemble, one can enhance the light-matter coupling. There was observed   entanglement
between two atomic ensembles located in spatially separated positions \cite{Chou05}, and between two spatial waves
of the collective atomic coherence in a single atomic ensemble  \cite{Yuan08}.
Through three-photon interference, the three cold atomic ensembles in independent quantum memory cells were heraldedly
entangled via measuring the photons and applying feedforward \cite{Jing19}.
Heralded single excitations can be created and stored as collective spin waves in a room temperature
atomic ensemble using spectral filtering performed by a high-finesse cavity \cite{Borregaard16}.
In general, the cavity-assisted schemes \cite{Borregaard16,Jing19} make it possible to effectively control
the light-matter coupling and the signal field mode structure, as well as the linewidth of the generated photon.

In the continuous variables (CV) domain \cite{Braunstein05a}, a  resource for squeezing and entanglement
is provided by the quantized light interaction with an atomic ensemble \cite{Raizen87,Cerf07,Hammerer10}.
If stimulated Raman scattering (SRS) is involved, multi-atomic collective spin in the medium supports amplification
of  spontaneous Stokes photons \cite{Raymer90,Sorensen09}. The CV squeezing and entanglement in the Raman schemes
were investigated in many works, including free space \cite{Wasilewski06} and cavity-assisted configurations
\cite{Sorensen02,Parkins06,Guzman06}.
An atom-light nonlinear interferometer \cite{Chen15} based on the interference of subsequent events of SRS in
an atomic ensemble is a sensitive probe of the atomic or optical phase change.

In this paper, we investigate theoretically the non-adiabatic effects in cavity-assisted Raman schemes for
generation of squeezing and entanglement, operated beyond the bad cavity limit in both the low-photon and CV regimes.
Starting from the bad cavity limit, we demonstrate how an increase in the cavity field lifetime results in better light-matter
coupling. On the other hand, we show that far beyond the bad cavity regime one can properly shape the signal only by
applying a stronger control field.  We reveal an optimal cavity--atomic ensemble matching with respect to the
cavity field lifetime, which allows one to apply less intense control field. Surprisingly, our optimal estimate for the
cavity field lifetime as compared to the signal duration is approximately the same for the low-photon and CV operation modes.

We find the classical control field time profiles that allow one to retrieve from the cavity the output quantized
signal of a predefined time shape and duration, which is optimal for the homodyne detection, optical mixing or
further manipulation. This is done  for a wide range of the signal duration as compared to the cavity field lifetime.
Our results make it possible to minimize a variety of non-linear effects, such as  AC light shifts,
four-wave mixing, etc, which may be potentially harmful to an experiment.

The collective spin-output signal entanglement and squeezing are analyzed by making use of the Bogolyubov
transformation and Bloch-Messiah reduction \cite{Braunstein05b}. In our model, this can be done in a general form, that is,
in terms of Green's functions.

Our approach  can be generalized to the generation of  multimode entangled states of light
and atoms. There is a variety of schemes of essentially multimode operation of the atom-light interaction in the low-photon
\cite{Kuhn15} and CV regime \cite{Vasilyev08,Chrapkiewicz12,Zhang13,Parigi15,Vetlugin16}
using  spatio-temporal patterns of both the classical control and the signal field, or a direct manipulation of the collective spin
waves by means of the spatially structured AC Stark shift \cite{Parniak19}, or by making use of the controllable
Zeeman shift \cite{Albrecht15} of atomic levels.

The suggested optimization of the generated signals duration, their temporal profile and degree of entanglement
and squeezing when applied to multimode regimes of operation may significantly speed-up quantum protocols leading
towards robust and efficient quantum communication, quantum computation and quantum metrology.
%%%%%%%%%%%%%%%%%%%%%%%%%%%%%%%%%%%%%%%%%%%%%%%%%
%%%%%%%%%%%%%%%%%%%%%%%%%%%%%%%%%%%%%%%%%%%%%%%%%
%%%%%%%%%%%%%%%%%%%%%%%%%%%%%%%%%%%%%%%%%%%%%%%%%
%%%%%%%%%%%%%%%%%%%%%%%%%%%%%%%%%%%%%%%%%%%%%%%%%

 \section{Cavity-assisted interaction of light with an atomic ensemble beyond the bad cavity limit}

\begin{figure}[h]
 \begin{center}
 \includegraphics[width=0.6\linewidth]{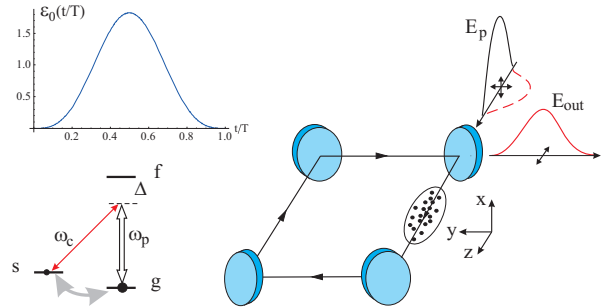}
 \caption{Schematic of cavity-assisted Raman system for generation of light-matter entanglement. A control field (double arrow)
induces two-quantum transition in $\Lambda$-scheme, and atomic collective spin excitation is stored in the coherence between the lower
levels (gray arrow). The output Stokes signal of duration $T$ is assumed to have predefined quasi-Gaussian time profile $\E_0(t/T)$,
see equation (\ref{E0_def}).}
 \label{figure1}
 \end{center}
 \end{figure}

The  scheme to be considered appears in figure \ref{figure1}.
%%%%%%%%%%%%%%%%%%%%%%%%%%%%%%%%%%%%%%%%%%%%%%%%%
%%%%%%%%%%%%%%%%%%%%%%%%%%%%%%%%%%%%%%%%%%%%%%%%%
The Hamiltonian of electric dipole interaction of $N$ motionless atoms with the cavity field in the rotating wave
approximation is given by $ H = H_0 + V$, where
 $$
 H_0 = \hbar \omega_{c}a^{\dagger}a + \hbar\sum_{j=1}^N
\left(\omega_{sg}\sigma^{(j)}_{ss}  + \omega_{fg}\sigma^{(j)}_{ff}\right),
 $$
 $$
 V = - \hbar \sum_{j=1}^N \left[\Omega (t)\sigma^{(j)}_{fg} e^{-i\omega_{p}t} +
a g\sigma^{(j)}_{fs}\right] + h.c.
 $$
Here $a$ is  quantized cavity field which we consider in single-mode approximation,
and $\sigma^{(j)}_{nm} = (|n\ra\la m|)^{(j)}$ is  the
atomic transition operator for the j-th atom, where $\omega_{nm}$ is the transition frequency.
The interaction Hamiltonian written in the Schrodinger picture explicitly depends on
classical control field with Rabi frequency (the slow amplitude) $\Omega(t)$.
The control field  Raman scattering produces bosonic quanta pairs (the light and the collective spin, see below).

Similar to parametric generation of pairs in $\chi^{(2)}$ non-linear media, in the limit of a weak coupling
this process can be considered as an analog of spontaneous parametric downconversion, while
strong coupling makes it possible to achieve, in the regime of parametric superluminescence,
a high degree of squeezing and entanglement between the light and matter degrees of freedom.

The cavity frequency $\omega_c$ and the classical control field frequency
$\omega_p = \omega_c + \omega_{sg}$ are matched in such a way as to support two-quantum resonance,
$g$ is coupling parameter for the quantized mode field.

Spatial factors are omitted in the Hamiltonian since for the co-propagating control and quantized fields the difference
between the longitu\-di\-nal wave numbers $k_{pz} - k_{cz}$ does not manifest itself on the atomic cloud length.
This can be a good approximation for atomic coherence generated within the same hyperfine level.
In case of  two different hyperfine levels, the spin polariton might be
of an essentially space-dependent form. We do not consider here  spatial addressability resource which allows for
an essentially multimode light-atoms interface operation \cite{Vetlugin16}.

The slow amplitudes of the field and the collective atomic observables are introduced as
 \BE
 \label{slow_ini_1}
 {\cal E}(t) = a(t)\exp(i\omega_c t),
 \EE
 \BE
 \label{slow_ini_2}
 \sigma_{gs}(t) = \sum_{j=1}^N \sigma^{(j)}_{gs}(t) e^{i\omega_{sg}t},
  \EE
 \BE
 \label{slow_ini_3}
\sigma_{sf}(t) = \sum_{j=1}^N \sigma^{(j)}_{sf}(t) e^{i\omega_c t}, \quad \sigma_{gf}(t) =
\sum_{j=1}^N \sigma^{(j)}_{gf}(t) e^{i(\omega_{c}+\omega_{sg})t}.
 \EE
The Heisenberg equations of motion are derived and linearized under the assumption that
the ground state g does not change its population within the evolution time interval $(0,T)$,
$\sigma^{(j)}_{gg}(t)\to 1$, while the population $\sigma^{(j)}_{ss}$ and $\sigma^{(j)}_{ff}$ can be neglected.
The cross-coherence $\sigma^{(j)}_{sf}$ is small in this limit, but must be accounted for since it
couples the cavity field to the atoms.

By introducing the cavity field decay at a rate $\kappa$ and the input cavity field ${\cal E}_{in}(t)$,
we arrive at
 \begin{equation}
 \label{dot_E_ini}
\dot{\cal E}(t) = -\kappa{\cal E}(t) + ig\sigma_{sf}(t)+ \sqrt{2\kappa}{\cal E}_{in}(t),
 \end{equation}
  \begin{equation}
 \label{dot_gs_ini}
 \dot\sigma_{gs}(t) =  -i\Omega(t)\sigma_{fs}(t) + ig{\cal  E}^{\dagger}(t)\sigma_{gf}(t),
 \end{equation}
 \begin{equation}
 \label{dot_gf_ini}
\dot{\sigma}_{gf}(t)= - i\Delta\sigma_{gf}(t) + i\Omega(t)N + ig{\cal E}(t)\sigma_{gs}(t),
\end{equation}
 \begin{equation}
 \label{dot_sf_ini}
\dot\sigma_{sf}(t) = - i\Delta\sigma_{sf}(t) + i\Omega(t)\sigma_{sg}(t),
\end{equation}
where $\Delta = \omega_{fg} - \omega_p$ is the Raman frequency mismatch.
The output quantized field amplitude is given by standard input-output relation,
 $$
{\cal E}_{out}(t)=\sqrt{2\kappa}\,{\cal E}(t)-{\cal E}_{in}(t),
 $$
which is valid for a high-finesse cavity with a close to unity reflectivity of the cavity mirrors.
The free-space input and output fields  satisfy standard commutation relations,
 $$
[{\cal E}_{in}(t),{\cal E}_{in}^\dagger(t')] = [{\cal E}_{out}(t),{\cal E}_{out}^\dagger(t')] = \delta(t-t'),
 $$
which preserve the commutation relation  $[{\cal E}(t),{\cal E}^{\dagger}(t)]=1$ for the cavity field.

Next, we  perform the adiabatic elimination.
That is, we assume the Raman regime condition when the frequency mismatch
$|\Delta|$ is much larger than other frequency parameters of the scheme.
The quantities  (d/dt)$\sigma_{gf}$ and (d/dt)$\sigma_{sf}$ are set to zero,
the corresponding cross-coherences are expressed in terms of other variables and substituted into the remaining
equations. In the following, we use the notation $S(t)$ for the collective spin
amplitude, $S(t) = \sigma_{gs}(t)/\sqrt{N}$. It is straightforward to demonstrate that due to the
property $\sigma_{kl}(t)\sigma_{mn}(t) = \delta_{lm}\delta_{kn}$ of the atomic operators, the
collective spin amplitude $S(t)$ is bosonic under the assumption of zero population of $s$ and $f$ states,
 $$
[S(t),S^\dag(t)] = 1.
 $$
We arrive at
 \begin{equation}
 \label{dot_E}
\dot{\cal E}(t) = -\kappa{\cal E}(t) + i
\frac{g\sqrt{N}}{\Delta}
\Omega(t){S}^{\dagger}(t)+ \sqrt{2\kappa}{\cal E}_{in}(t),
 \end{equation}
 \begin{equation}
\label{dot_S}
\dot{{S}}(t) = - i\left[\frac{|\Omega(t)|^2}{\Delta} - g^2\frac{\E^\dag\E}{\Delta}\right]S(t)
+  i\frac{g\sqrt{N}}{\Delta}\Omega(t){\cal E}^{\dagger}(t) .
 \end{equation}
Now we introduce physically reasonable corrections to the observables' frequencies.
The cavity mode frequency shift due to linear refractive index is absent in our model,
since the lower state $s$ of the $f - s$ transition is not populated.
The dynamic correction $\delta_s(t)$ to the frequency $\omega_{sg}$ of the  $s - g$ transition
due to AC Stark shift of the ground state, induced by the strong control field,  results in an additional
phase $\varphi_s(t)$ of the collective spin coherence, $\sigma_{gs}(t) \sim \exp(-i\varphi_s(t))$, where
 $$
 \delta_s(t) = \frac{|\Omega(t)|^2}{\Delta}, \qquad  \varphi_s(t) = \int_0^t dt' \delta_s(t').
 $$
The frequency shift of the state $s$ induced by the quantized cavity field will be dropped, since the
number of cavity photons is assumed to be relatively small, $\la\E^\dag\E \ra \ll |\Omega|^2$.
The phase correction $\varphi_s(t)$ is incorporated into a new self-consistent set of slow field and atomic variables,
 $$
 S(t) = e^{-i\varphi_s(t)} \tilde S(t), \qquad   \Omega(t) = e^{-i\varphi_s(t)} \tilde\Omega(t).
 $$
This yields,
 \begin{equation}
 \label{dot_E_corr}
\dot{\cal E}(t) = -\k{\cal E}(t) + i\frac{g\sqrt{N}}{\Delta}\tilde\Omega(t){\tilde S}^{\dagger}(t) +
\sqrt{2\kappa}{\cal E}_{in}(t),
 \end{equation}
 \begin{equation}
\label{dot_S_corr}
\dot{\tilde S}(t) =  i\frac{g\sqrt{N}}{\Delta}{\tilde\Omega}(t){\cal E}^\dag(t).
 \end{equation}
In the following, we drop the tildes in these equations for brevity.

Let us introduce dimensionless time $\tau = 2\kappa t$ measured in the units of the cavity field lifetime.
The equation (\ref{dot_E_corr}) together with the Hermitian conjugate counterpart of (\ref{dot_S_corr})
make up a complete set of linear equations for the field and spin observables,
 \begin{equation}
 \label{dot_E_diml}
\frac{d}{d\t}\E(\t) = -\frac{1}{2}\E(\t) + ik(\t)S^\dag(\t) + \E_{in}(\t),
 \end{equation}
 \begin{equation}
\label{dot_Sdag_diml}
\frac{d}{d\t}S^\dag(\t) = -  ik^*(\t)\E(\t).
 \end{equation}
Here
 \BE
 \label{coupling}
 k(\t) = \frac{g\sqrt{N}}{2\k\Delta}\Omega(\t),
 \EE
is dimensionless coupling parameter, proportional to the control field strength.
The dimensionless free-space fields, $\E_{in}(t)/\sqrt{2\k} \to \E_{in}(\t)$, satisfy the commutation relation
 $$
[\E_{in}(\t),\E_{in}^\dag(\t')] = [\E_{out}(\t),\E_{out}^\dag(\t')] = \delta(\t - \t').
 $$
One can represent the solution of linear equations (\ref{dot_E_diml}, \ref{dot_Sdag_diml}) for $\E$ and $S^\dag$
in terms of dimensionless Green's functions $G_{nm} (\t,\t')$, $n, m = \E, S^\dag$.  For the observables
$S$, $\E$ and $\E^{(out)}$ this yields:
 \BE
 \label{sol_S}
S(\t) = G^*_{S^\dag S^\dag}(\t,0)S(0)  + G^*_{S^\dag \E}(\t,0)\E^\dag(0)  +
\int_0^\t d\t' G^*_{S^\dag \E}(\t,\t')\E^\dag_{in}(\t'),
 \EE
 \BE
 \label{sol_E}
\E(\t) = G_{\E\E}(\t,0)\E(0)   + \int_0^\t d\t' G_{\E\E}(\t,\t')\E_{in}(\t') +
G_{\E S^\dag}(\t,0)S^\dag(0),
 \EE
 \begin{equation}
\label{in_out_diml}
{\E}_{out}(\t) = {\E}(\t) - {\E}_{in}(\t).
 \end{equation}
The equations (\ref{sol_S} - \ref{in_out_diml}) represent the Bogolyubov transformation,
which for a general set of input-output field amplitudes, $\vec a^{(in)} \sim \{a_n^{(in)}\}$,
$\vec a^{(out)} \sim \{a_n^{(out)}\}$, is given by
 \BE
 \vec a^{(out)} = A \vec a^{(in)} + B \vec a^{(in)\dag}.
 \label{Bogol}
 \EE
In terms of the Bloch-Messiah reduction \cite{Braunstein05b}, the matrices $A$ and $B$ are represented
as
 \BE
 A = UA^{(D)}V^\dagger, \qquad B = UB^{(D)}V^T,
 \label{BMred}
 \EE
where $U$ and $V$ are unitary matrices, and $A^{(D)}$, $B^{(D)}$ -- non-negative (that is, with
non-negative eigen numbers) diagonal matrices, such that
 $$
 A^{(D)2} - B^{(D)2} = I.
 $$
The bosonic input and output amplitudes of the modes that are eigen for the Bogolyubov transform are
 $$
 \vec I = V^\dag \vec a^{(in)}, \quad \vec O = U^\dag \vec a^{(out)},
 $$
 or
 $$
 I_n = \sum_m V_{mn}^* a_m^{(in)}, \qquad O_n = \sum_m U_{mn}^*a_m^{(out)}.
 $$
Inserting these definitions together with (\ref{BMred}) into (\ref{Bogol}) one arrives at
squeezing transformations of the form
 \BE
 \label{BM_diag}
O_n = A_n^{(D)} I_n + B_n^{(D)} I^\dag_n.
 \EE
The sets of input and output quantized amplitudes which undergo squeezing correspond to non-zero values $B_n^{(D)}$.

In our model, one can find the Bloch-Messiah representation of the field and collective spin evolution
in a general form (that is, in terms of Green's functions), see Appendix. There are two input and two output
bosonic amplitudes that undergo degenerate squeezing in agreement with (\ref{AD_BD}). These amplitudes
are found to be
 \BE
 \label{in}
 I_1 = \frac{e^{-i\xi_G/2}}{\sqrt{2}}\left(I'_1 + I'_2\right), \quad
 I_2 = -i\frac{e^{-i\xi_G/2}}{\sqrt{2}}\left(I'_1 - I'_2\right),
 \EE
and
 \BE
 \label{out}
 O_1 = \frac{e^{i\xi_G/2}}{\sqrt{2}}\left(O'_1 + O'_2\right), \quad
 O_2 = i\frac{e^{i\xi_G/2}}{\sqrt{2}}\left(O'_1 - O'_2\right),
 \EE
respectively, where $I'_{1,2}$ and $O'_{1,2}$ are defined as
 \BE
 \label{in1'}
 I'_1 = \frac{1}{\sqrt{N_1}} \left(G_{S^\dag \E}(\T,0)\E(0)  +
 \int_0^\T d\t G_{S^\dag \E}(\T,\t)\E_{in}(\t)\right),
 \EE
 \BE
 \label{in2'}
 I'_2 = S(0),
 \EE
and
 \BE
 \label{out1'}
 O'_1 = S(\T),
 \EE
 \BE
 \label{out2'}
O'_2 =  \frac{1}{\sqrt{N_2}} \left(G^*_{\E S^\dag}(\T,0)\E(\T)  +
\int_0^\T d\t G^*_{\E S^\dag}(\t,0)\E_{out}(\t)\right).
 \EE
Here  $G_{S^\dag S^\dag}(\T,0) = e^{i\xi_G/2}\left|G_{S^\dag S^\dag}(\T,0)\right|$,
and  $N_1$, $N_2$ are the normalization coefficients,
 $$
 N_1 = |G_{S^\dag \E}(\T,0)|^2  +  \int_0^\T d\t |G_{S^\dag \E}(\T,\t)|^2, \quad
 N_2 = |G_{\E S^\dag}(\T,0)|^2  + \int_0^\T d\t |G_{\E S^\dag}(\t,0)|^2,
 $$
 such that
 $$
 [I'_n,I'^\dag_m] = [O'_n,O'^\dag_m] = \delta_{nm}.
 $$
As shown in Appendix (see (\ref{N12})),
 $$
 N_1 = N_2 = |G_{S^\dag S^\dag}(\T,0)|^2 - 1.
 $$
In the following, we shall discuss the shape of the control field $\Omega(\t)$ which allows one to retrieve
an output signal in a given temporal mode $\E_0(\t)$, convenient for the detection or further manipulation.
This mode is assumed to have a normalized quasi-Gaussian shape of duration $\T$, shown in figure
\ref{figure1},
 \BE
 \label{E0_def}
 \E_0(\t) = N_\E \left[\exp[-4(\t/\T - 1/2)^2] - e^{-1}\right]\cos^2[\pi(\t/\T -1/2)],
 \EE
 $$
 \int_0^\T d\t \E_0^2(\t) = 1,
 $$
where $N_\E$ is the normalization coefficient. Quantized amplitude $\E_d$ of the signal  is given
by the projection of the output field onto $\E_0$,
 $$
 \E_d = \int_0^\T d\t \E^*_0(\t) \E_{out}(\t), \quad [\E_d, \E_d^\dag] = 1.
 $$
We assume that one controls the system in such a way that
 \BE
 \label{E0_G}
\frac{1}{\sqrt{N_0}}G_{\E S^\dag}(\t,0) = \E_0(\t),
 \EE
where $N_0$ is the normalization coefficient. It is evident from (\ref{out2'}) that  the output amplitude $O'_2$
is a superposition of the signal amplitude $\E_d$ and the resulting cavity field,
 $$
 O'_2 = \frac{1}{\sqrt{N_2}}\left(G^*_{\E S^\dag}(\T,0) \E(\T) + \sqrt{N_0}\E_d \right).
 $$
Here all bosonic amplitudes are properly normalized if $N_0 =  |G_{S^\dag S^\dag}(\T,0)|^2 -
|G_{\E S^\dag}(\T,0)|^2  - 1$.
In general, the output signal amplitude $\E_d$ is not eigen for the squeezing transformation and is
represented by a beamsplitter-like equation
 $$
 \E_d = \sqrt{\eta} O'_2 + \sqrt{1 - \eta} O'_v.
 $$
The amplitude $O'_v$ represents an output oscillator which does not undergo squeezing
(that is, belongs to a set of degrees of freedom for which $B^{(D)}_j = 0$, $j=v$), and is assumed to be in vacuum state.
Here $\sqrt{\eta}$ is projection of the signal mode (\ref{E0_G}) onto the eigen output mode $O'_2$, and
 \BE
 \label{eta}
\eta = \frac{N_0}{N_2} = 1 - \frac{|G_{\E S^\dag}(\T,0)|^2}{ |G_{S^\dag S^\dag}(\T,0)|^2 - 1} \leq 1,
 \EE
is quantum efficiency of the retrieval of photons generated by  Raman luminescence by
the time $\T$ (by the end of observation)  from the cavity.

Finally, we make use of (\ref{out}) and relate the observables $S(\T)$ and $\E_d$ to the output
eigenmodes of squeezing,
 \BE
 S(\T) = \frac{e^{-i\xi_G/2}}{\sqrt{2}}(O_1 - iO_2),
 \EE
 \BE
 \label{Ed}
 \E_d = \frac{e^{-i\xi_G/2}}{\sqrt{2}}\sqrt{\eta}(O_1 + iO_2) + \sqrt{1 - \eta}O_v.
 \EE
Quadrature amplitudes of the observables $S(\T)$ and $\E_d$ that make it possible to minimize the
variance used in the Duan inseparability criterion \cite{Duan00} for continuous variables are introduced as
 $$
 X_S = {\rm Re}\, e^{i\xi_G/2} S(\T), \qquad Y_S = {\rm Im}\, e^{i\xi_G/2} S(\T),
 $$
and similar for $X_\E$, $Y_\E$. For vacuum initial conditions the variance itself is found to be
 $$
 \la \left[(X_S - X_\E)^2 + (Y_S + Y_\E)^2 \right]\ra =
 $$
 \BE
 \label{Duan}
\frac{1}{4}\left[(\sqrt{\eta}-1)^2 e^{2r}  + (\sqrt{\eta}+1)^2 e^{-2r} + 2(1-\eta)\right].
 \EE
The stretching and squeezing factor $e^{\pm r}$ in terms of the Green's functions is given by (\ref{e_r}).
The retrieved signal mode $\E_0(\t)$ is assumed to be of the form (\ref{E0_def}) in our theory, which
implies $\E_0(\T) = 0$ and hence $G_{\E S^\dag}(\T,0) = 0$, see (\ref{E0_G}).  The signal retrieval
efficiency (\ref{eta}) is found to be $\eta = 1$, that is, no field left in the cavity by $\T$. The variance
(\ref{Duan}) is minimal in this limit and reads as
 $$
\la \left[(X_S - X_\E)^2 + (Y_S + Y_\E)^2 \right]\ra = e^{-2r}.
 $$
 %
%%%%%%%%%%%%%%%%%%%%%%%%%%%%%%%%%%%%%%%%%%%%%%%%%
%%%%%%%%%%%%%%%%%%%%%%%%%%%%%%%%%%%%%%%%%%%%%%%%%
%%%%%%%%%%%%%%%%%%%%%%%%%%%%%%%%%%%%%%%%%%%%%%%%%
%%%%%%%%%%%%%%%%%%%%%%%%%%%%%%%%%%%%%%%%%%%%%%%%%

 \section{Optimal control of squeezing and light-matter entanglement beyond the adiabatic limit}
 \label{sec_control}

In this section, we start by estimating the control field time profile that matches the predefined signal
mode (\ref{E0_def}) for different cavity decay rates, including the bad cavity limit as well as
essentially non-adiabatic regimes of the atoms--field interaction.

In order to reveal an optimal matching of the cavity field lifetime with the signal duration,
we further address the situation which might be of interest for an experiment. That is,
if there is a given atomic cell and a control field source, which cavity quality
has to be chosen in order to achieve a given degree of squeezing and entanglement for the signal mode of given
shape and duration by applying less intensive control field. This might be important
in order to minimize a variety of harmful non-linear effects that might arise in the scheme.

The signal time profile is related to the Green's function $G_{\E S^\dag}$ by (\ref{E0_G}).
This Green's function can be found
from the semiclassical version of the basic equations (\ref{dot_E_diml}, \ref{dot_Sdag_diml}) with
the initial condition $\E(0) = 0$, $S^*(0) =1$.

The matter and the cavity field excitations are generated in pairs, and then the field
excitations leak through the output port. This leads to the semiclassical excitation balance of the form
 $$
 \frac{d}{d\t}(|S(\t)|^2 - |\E(\t)|^2) = |\E(\t)|^2,
 $$
as it follows from (\ref{dot_E_diml}, \ref{dot_Sdag_diml}).
Consider a signal with the normalized shape $\E_0(\t)$ and an arbitrary amplitude,
 \BE
 \label{signal}
\E(\t) = \sqrt{n_0}\E_0(\t).
 \EE
The time-dependent collective spin amplitude which matches the signal can be derived from
 $$
 |S(\t)|^2 = |S(0)|^2  + n_0\left[|\E_0(\t)|^2 - |\E_0(0)|^2 + \int_0^\t d\t' |\E_0(\t')|^2\right].
 $$
Assuming $S(\t)$ is real, the coupling parameter (\ref{coupling}) that matches (\ref{dot_E_diml})
is found to be
 \BE
 \label{coupling_ES}
 ik(\t) = \frac{\sqrt{n_0}}{|S(\t)| }\left[\frac{d}{d\t}\E_0(\t) + \E_0(\t)/2\right].
 \EE
We have chosen the signal mode profile $\E_0(\t)$ such that $\E(0) = 0$, and for the initial
condition $S^*(0) =1$ this just gives the control field shape which matches the needed form of the
Green's function $G_{\E S^\dag}(\t)$. Note that the semiclassical version of (\ref{sol_E}), where
$\E(0) = 0$ and $\E_{in}(\t)\to 0$, reads as $\E(\t) = G_{\E S^\dag}(\t)$.
It follows  from (\ref{E0_G}), that the norm of the signal
(\ref{signal}) is
 $$
 n_0 = N_0 = |G_{S^\dag S^\dag}(\T,0)|^2 - |G_{\E S^\dag}(\T,0)|^2  - 1.
 $$
As seen from (\ref{BM_diag}), the real and imaginary quadrature amplitudes of the output eigenmodes $O_n$
are stretched and squeezed with respect to these of the input eigenmodes $I_n$ by the factors $e^{r}$ and $e^{-r}$
respectively, where
 \BE
 \label{e_r}
 e^{\pm r} = A_n^{(D)} \pm B_n^{(D)} = |G_{S^\dag S^\dag}(\T,0)| \pm
 \sqrt{ |G_{S^\dag S^\dag}(\T,0)|^2 -1}.
 \EE
Since the output signal amplitude is related to the eigenmodes of squeezing by (\ref{Ed}), the average
number of photons
in the signal is found to be
 $$
 \la \E_d^\dag \E_d\ra = \frac{\eta}{2} [{\rm cosh}(2r) -1] = n_0.
 $$
In our calculations, the semiclassical version of (\ref{dot_E_diml}, \ref{dot_Sdag_diml}) was solved
numerically with the coupling parameter (\ref{coupling_ES}) and with the  initial conditions that match the Green's
functions $G_{S^\dag S^\dag}(\T,0)$ and $G_{\E S^\dag}(\t,0)$. The output signal mode profile found from
this procedure was in perfect agreement with the predefined one, see (\ref{E0_def}).

\begin{figure}[h]
 \begin{center}
 \includegraphics[width=0.6\linewidth]{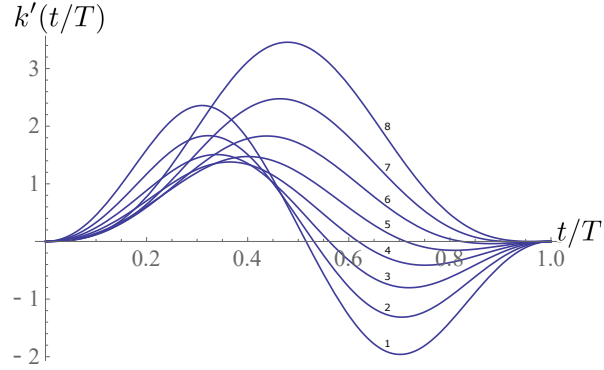}
 \caption{Control field time profiles defined in section \ref{sec_control}, evaluated for different values of the ratio
$T/t_c = 2\kappa T$ of the signal duration $T$ (which is assumed to be fixed in this calculation, while $t_c$ varies)
to the cavity field lifetime. The curves 1\ldots 8
correspond to $T/t_c$ = 1; 2; 4; 8; 16; 32; 64; 128, where the first curve represents an essentially non-adiabatic regime,
and the last one illustrates the bad cavity limit. In this plots the squeezing and entanglement parameters are $e^r = 1.4$
and $n_0 \approx 0.12$,  and correspond to spontaneous generation of random light-matter quanta pairs.}
 \label{figure2}
 \end{center}
 \end{figure}

\begin{figure}[h]
 \begin{center}
 \includegraphics[width=0.6\linewidth]{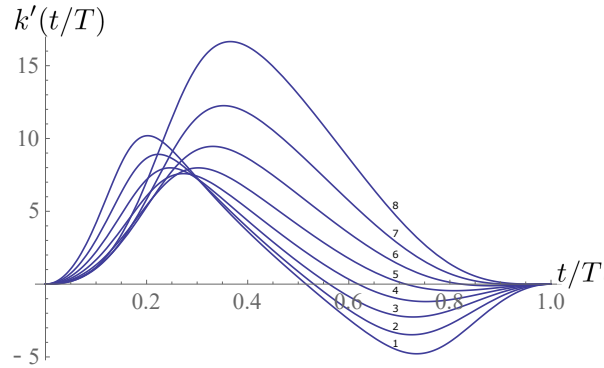}
 \caption{The same as in figure \ref{figure2} for medium degree of squeezing and entanglement,
 $e^r = 6.5$ and $n_0 \approx 10$.}
 \label{figure3}
 \end{center}
 \end{figure}

\begin{figure}[h]
 \begin{center}
 \includegraphics[width=0.6\linewidth]{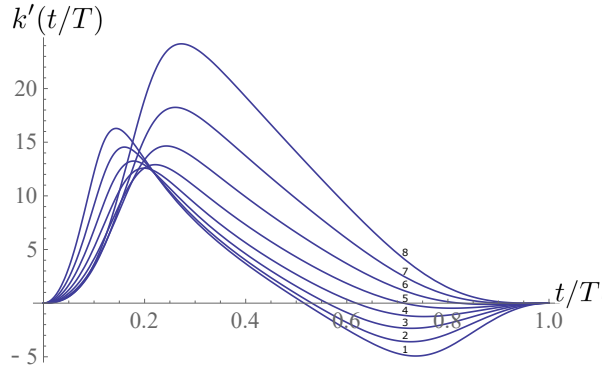}
 \caption{The same as in figure \ref{figure2} for large degree of squeezing and entanglement,
 $e^r = 20$ and $n_0 \approx 100$.}
 \label{figure4}
 \end{center}
 \end{figure}

Assuming the signal is of the given form and duration $T$ (in seconds), and the atomic cell parameters are
also given, we introduce the atoms--field coupling parameter $q(\t)$ whose definition does not change with
the cavity decay rate and represents the control field strength in physical units, multiplied by a constant factor,
 $$
 q(\t) = 2\k T k(\t) = T\frac{g\sqrt{N}}{\Delta}\Omega(\t).
 $$
We show in figures \ref{figure2} - \ref{figure4} the coupling parameter as a function of
the dimensionless time $t/T$. Namely, we plot the quantity $k'(t/T) \equiv q\big(\t = (t/T)\T\big)$
for a set of factors $e^r$ and corresponding average photon numbers $n_0$ in the retrieved signal.
%%%%%%%%%%%%%%%%%%%%%%%%%%%%%%%%%%%%%%%%%%%%%%%%%
%%%%%%%%%%%%%%%%%%%%%%%%%%%%%%%%%%%%%%%%%%%%%%%%%
%%%%%%%%%%%%%%%%%%%%%%%%%%%%%%%%%%%%%%%%%%%%%%%%%
%%%%%%%%%%%%%%%%%%%%%%%%%%%%%%%%%%%%%%%%%%%%%%%%%
The values $e^r = 1.4$ and $n_0 \approx 0.12$ represent spontaneous parametric generation of random
light-matter quanta pairs, which is commonly used for the probabilistic heralded creation of single collective spin
excitations. A medium and large degree of squeezing and entanglement are represented by
the values $e^r = 5$, $n_0 \approx 10$, and $e^r = 20$, $n_0 \approx 100$, respectively.

In the bad cavity limit,  the cavity excitations lifetime $t_c = 1/2\k$ is much less than
the signal duration, $T/t_c = 2\k T \gg 1$. Beyond the bad cavity limit, an essentially non-adiabatic regime
takes place  when the cavity field lifetime becomes comparable with the signal duration $T$.
As seen from the plots, in the latter case one needs control field shape which is switching its sign at
some time moment $t_s$.

The latter effect has an analogy in cavity assisted atomic quantum memories beyond
the bad cavity limit \cite{Veselkova17,Veselkova19}. Both in the memory and the squeezing configurations,
the cavity with a certain field lifetime does not allow to shape properly  a steep rear slope of the signal field
leaking out of the cavity after $t_s$. In both cases one has to switch at $t_s$ the direction of the energy flow,
and to convert some cavity photons back to the collective spin (memory), or to up-convert some pairs
of light and matter quanta back to the control field (squeezing). It is clear from this interpretation,
that the time moment $t_s$ is the same for both devices given the same cavity excitations lifetime and the
same signal shape, and does not depend on the degree of squeezing.
%%%%%%%%%%%%%%%%%%%%%%%%%%%%%%%%%%%%%%%%%%%%%%%%%
  \begin{figure}[h]
 \begin{center}
 \includegraphics[width=0.6\linewidth]{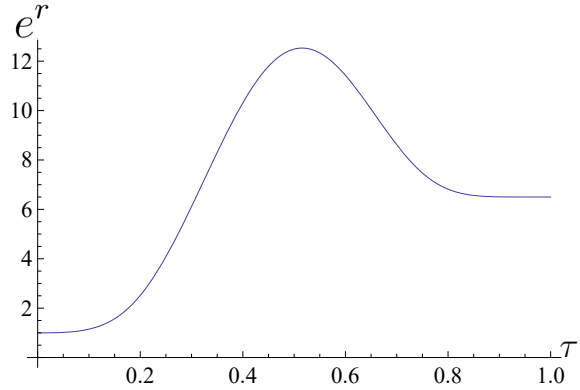}
 \caption{Time-dependent parameter of the squeezing and entanglement present in the system, evaluated for short
signal and medium degree of output entanglement, $T/t_c = 1$, $e^r = 6.5$ and $n_0 \approx 10$.}
 \label{figure5}
 \end{center}
 \end{figure}
%%%%%%%%%%%%%%%%%%%%%%%%%%%%%%%%%%%%%%%%%%%%%%%%%
In case of squeezing, the degree of squeezing and light-matter entanglement
achieved by the time moment $t_s$ decreases during  final stage of evolution, as illustrated in figure \ref{figure5}
for short signal, $T/t_c = 1$ and medium degree of the output squeezing and entanglement,
$e^r = 6.5$, and $n_0 \approx 10$. In this plot we represent time-dependent intermediate degree of squeezing
with the parameter $\exp(r(\t))$ evaluated using (\ref{e_r}), where $\T \to \t$.

An important distinction between the memory and the squeezing cavity-assisted schemes beyond the adiabatic
approximation is that for some applications the memories with essentially degraded quantum efficiency are considered
to be useless, but this is not the case for squeezing. Even for ``short'' signals in the timescale
of the cavity decay time one can achieve a needed degree of the output squeezing and entanglement
at the expense of the control field power, as seen from the plots \ref{figure2} -
\ref{figure4} for $T/t_c$ = 8; 4; 2; 1.

Starting from the bad cavity limit, $T/t_c$ = 128; 64; 32; 16; 8, one can achieve the same degree
of squeezing and entanglement by increasing the cavity field lifetime and applying less intense control field,
since this provides a longer interaction time.

Surprisingly, an optimal matching of the cavity excitations lifetime with the signal duration, which assures
a minimal control field peak power, is achieved for $T \approx 8 t_c$ for any degree of squeezing and entanglement.

%%%%%%%%%%%%%%%%%%%%%%%%%%%%%%%%%%%%%%%%%%%%%%%%%
%%%%%%%%%%%%%%%%%%%%%%%%%%%%%%%%%%%%%%%%%%%%%%%%%
%%%%%%%%%%%%%%%%%%%%%%%%%%%%%%%%%%%%%%%%%%%%%%%%%
%%%%%%%%%%%%%%%%%%%%%%%%%%%%%%%%%%%%%%%%%%%%%%%%%

 \appendix

 \section{Eigenmodes of squeezing and Bloch-Messiah reduction}

We consider the field and collective spin evolution within a given time interval $(0,T)$. The complete
initial set of bosonic amplitudes at $t = 0$ is composed of $S(0)$, $\E(0)$, and $\{\E_{in}(\t)\}$,
where $\t$ runs from 0 to $\T = 2\k T$. Note that the field $\E_{in}(\t)$ arrives at the input port at a time
moment $\t$, and initially is located at a distance - $ct$ from the cavity. Hence, the amplitudes $\E_{in}(\t)$
can be viewed at as initial field amplitudes at remote locations in the Heisenberg picture.

Similarly, the complete final set of amplitudes at $t=T$ is composed of $S(\T)$, $\E(\T)$, and $\{\E_{out}(\t)\}$,
where $ 0<\t<\T$, and the Bogolyubov transformation  (\ref{sol_S} - \ref{in_out_diml}) is equivalent to unitary evolution
on the time interval $(0,\T)$, where bosonic commutation relations of the observables are preserved if
 \BE
 \label{ABprop}
AA^\dagger - BB^\dagger = I, \qquad AB^T = BA^T.
  \EE
In order to  diagonalize $B$, one has to single out the terms that perform mapping
$a_n^{(in)\dag} \to a_m^{(out)}$ in (\ref{sol_S} - \ref{in_out_diml})  and hence belong to $B$.
The relevant input and output bosonic amplitudes were already introduced in (\ref{in1'} - \ref{out2'}).
The matrices of the Bogolyubov tranformation (\ref{sol_S} - \ref{in_out_diml}) that arise in the basis
of the introduced above bosonic amplitudes are labeled $A'$, $B'$, etc.

Let us explain the structure of the amplitudes (\ref{in1'} - \ref{out2'}).
The conjugates of the input  amplitudes $I'_n$, $n=1,2$, are explicitly present in the right side of
(\ref{sol_S} - \ref{in_out_diml}), so the mapping attributed to $B$ is given by
 \BE
 \label{B_terms1}
 S^B(\T) = \sqrt{N_1}I'^\dag_1, \quad \E^B(\T) =  G_{\E S^\dag} (\T,0) I'^\dag_2,
 \EE
 \BE
 \label{B_terms2}
 \E^B(\t) =  G_{\E S^\dag} (\t,0)I'^\dag_2,
 \EE
where the superscript $B$ specifies the relevant contributions.
An arbitrary output bosonic amplitude $O'_m$ of the form
 $$
 O'_m = C^{(m)}_{S}S(\T) + C^{(m)}_{\E}\E(\T) +
 \int_0^\T d\t C^{(m)}_{\E}(\t)\E(\t),
 $$
is specified by a normalized complex vector
$\{C^{(m)}_{S}, C^{(m)}_{\E}, \{C^{(m)}_{\E}(\t)\}\}$.
The output amplitudes $O'_n$  (\ref{out1'}, \ref{out2'}) for $n=1,2$, are represented by the vectors
$(1/\sqrt{N_2})\{0,G^*_{\E S^\dag}(\T),\{G^*_{\E S^\dag}\}\}$ and $\{1,0,0\}$ respectively.

The negative-frequency contributions to $O_n'$, $n=1,2$,  that stem from $B$ are evaluated by
using (\ref{B_terms1}, \ref{B_terms2}) and arise in the form
 $$
 O_1'^B = \sqrt{N_1} I_1'^\dag, \qquad O_2'^B = \sqrt{N_2} I_2'^\dag.
 $$
It follows from (\ref{B_terms1}, \ref{B_terms2}) that if the vector which represents $O'_m$ is orthogonal
to these for $O'_1$ and $O'_2$, the amplitude $O'_m$ does not undergo squeezing, that is, $O_m'^B =0$.
This yields diagonal representation of $B'$:
 \BE
 \label{B_mapping}
 \vec O'^B = B' \vec I'^{\dag},  \qquad O'^B_n = \sum_n B'_{nm} I'^\dag_m, \quad n,m = 1,2,
 \EE
where $B'_{11} = B'^{(D)}_1  = \sqrt{N_1}$, $B'_{22} = B'^{(D)}_2 = \sqrt{N_2}$.

In the case of degenerate squeezing, this diagonal representation still does not ensure
diagonal form of $A'$. We have to admit for the introduced above input and output amplitudes
which undergo squeezing transformation that the mapping performed by $A$ is
 \BE
 \label{A_mapping}
\vec O'^A = A' \vec I', \qquad  O'^A_n = \sum_n A'_{nm} I'_m, \quad n,m = 1,2.
 \EE
Here  the contributions to the output amplitudes that are due to $A$ are labeled $O'^A_n$.
It follows from (\ref{sol_S} - \ref{in_out_diml}), that (i) $O'^A_1 = G_{S^\dag S^\dag}^*(\T,0) I'_2$,
and (ii) $A$ performs mapping between the input and the output field degrees of freedom only.
The latter implies that  there is no other possibility for $A$ as to map $I'_1$ onto $O'_2$ within
the essential subsets of oscillators specified by (\ref{in1'} - \ref{out2'}).
The non-zero matrix elements of $A'$ are $A'_{12}  = G_{S^\dag S^\dag}^*(\T,0)$ and $A'_{21}$.

It follows from the first equation (\ref{ABprop}) that
 \BE
 \label{N12}
 N_1 = |G_{S^\dag S^\dag}(\T,0)|^2 - 1, \qquad N_2 = |A'_{21}|^2 - 1,
 \EE
and, consequently,  the second one yields $A'_{21} = G_{S^\dag S^\dag}^*(\T,0)$ and $N_2 = N_1$.
We arrive at the Bogolyubov transformation $\vec O' = A'\vec I' + B' \vec  I'^\dag$, where
 $$
 A' = G_{S^\dag S^\dag}^*(\T,0)\left[\BA{cc} 0 & 1\\ 1 & 0 \EA\right], \quad
 B' = \sqrt{|G_{S^\dag S^\dag}(\T,0)|^2 - 1}\left[\BA{cc} 1 & 0\\ 0 & 1 \EA\right].
 $$
In order to bring $A'$ to diagonal form, we define new input and output bosonic amplitudes
within the subsets of observables that undergo degenerate squeezing,
 \BE
 \label{in_in'}
  \vec I = D^* \vec I', \quad \vec O = D \vec O',
 \EE
where $D$ is a unitary matrix. Similar to (\ref{B_mapping}, \ref{A_mapping}), new Bogolyubov transformation
matrices $A$ and $B$ are introduced via
 $$
\vec O^A =  A \vec I, \qquad \vec O^B =  B \vec I^\dag.
 $$
This yields
  \BE
  \label{ABnew}
 A = DA'D^T, \qquad B = DB'D^\dag = B'.
  \EE
Let us define $D$ as
 $$
 D = \frac{e^{i\xi_G/2}}{\sqrt{2}} \left[\BA{rc} 1 & 1\\ i & -i \EA \right],
 $$
where $G_{S^\dag S^\dag}(\T,0) = e^{i\xi_G/2}\left|G_{S^\dag S^\dag}(\T,0)\right|$.
Inserting this into (\ref{ABnew}), we arrive at the diagonal form of $A$ and $B$ and eventually at the
Bloch-Messiah representation (\ref{BM_diag}) for our model, where
 \BE
 \label{AD_BD}
 A^{(D)}_{1,2} = \left|G_{S^\dag S^\dag}(\T,0)\right|, \qquad
 B^{(D)}_{1,2} = \sqrt{|G_{S^\dag S^\dag}(\T,0)|^2 - 1},
 \EE
and the input and output eigenmodes of squeezing are given by (\ref{in_in'}).

%%%%%%%%%%%%%%%%%%%%%%%%%%%%%%%%%%%%%%%%%%%%%%%%%
%%%%%%%%%%%%%%%%%%%%%%%%%%%%%%%%%%%%%%%%%%%%%%%%%
%%%%%%%%%%%%%%%%%%%%%%%%%%%%%%%%%%%%%%%%%%%%%%%%%
%%%%%%%%%%%%%%%%%%%%%%%%%%%%%%%%%%%%%%%%%%%%%%%%%

 \section*{Conclusion}

We have considered the non-adiabatic effects in quantum entanglement between light and collective spin polarization
of a cold atomic ensemble. Entanglement we deal with results from the interaction of Stokes light wave with atoms
in cavity-assisted scheme both in the low-photon and CV regimes. We assume that the retrieved Stokes signal has a
predefined time shape suitable for optical mixing and homodyne detection, which is important for quantum repeaters,
entanglement swapping, one-way quantum computation etc. 
In order to increase light-matter coupling, it is reasonable to use cavities with larger
cavity field lifetime, while use of shorter signals may improve the scheme operation speed and overcome the effect of
atomic relaxation. We have investigated the non-adiabatic effects that arise beyond the bad cavity limit and suggested
a physical explanation of these effects. We find the control field time profiles that result in the retrieval of the signal of
a given time shape and duration for different cavity field lifetimes. An optimal ratio of the signal duration to the
cavity field lifetime that makes it possible to apply control field with less peak intensity and to minimize a variety of
potentially harmful non-linear effects is revealed.

%%%%%%%%%%%%%%%%%%%%%%%%%%%%%%%%%%%%%%%%%%%%%%%%%
%%%%%%%%%%%%%%%%%%%%%%%%%%%%%%%%%%%%%%%%%%%%%%%%%
%%%%%%%%%%%%%%%%%%%%%%%%%%%%%%%%%%%%%%%%%%%%%%%%%
%%%%%%%%%%%%%%%%%%%%%%%%%%%%%%%%%%%%%%%%%%%%%%%%%

 \section*{Acknowledgments}

This research was supported by the Russian Foundation for Basic Research (RFBR) under the projects
19-02-00204-a and 18-02-00648-a. NIM acknowledges the RFBR grant for young researchers 18-32-00255-mol-a.

%%%%%%%%%%%%%%%%%%%%%%%%%%%%%%%%%%%%%%%%%%%%%%%%%
%%%%%%%%%%%%%%%%%%%%%%%%%%%%%%%%%%%%%%%%%%%%%%%%%
%%%%%%%%%%%%%%%%%%%%%%%%%%%%%%%%%%%%%%%%%%%%%%%%%
%%%%%%%%%%%%%%%%%%%%%%%%%%%%%%%%%%%%%%%%%%%%%%%%%

 %\newpage

 \section*{References}
 %\bibliography{}
 %\bibliographystyle{plain}

\end{document}